\documentclass{article}

\usepackage{moreverb,url}
\usepackage{amsmath}
\usepackage{amssymb}
\usepackage{subfig}
\usepackage{color}
\usepackage{enumitem}
\usepackage{subfig}
\usepackage{tikz}
\usepackage{natbib}
\usepackage{booktabs}
\usepackage[affil-it]{authblk} 
\usetikzlibrary{shapes,decorations,arrows,calc,arrows.meta,fit,positioning}
\usepackage[margin=1.2in]{geometry}
\tikzset{
    -Latex,auto,node distance =1 cm and 1 cm,semithick,
    state/.style ={ellipse, draw, minimum width = 0.7 cm},
    point/.style = {circle, draw, inner sep=0.04cm,fill,node contents={}},
    bidirected/.style={Latex-Latex,dashed},
    el/.style = {inner sep=2pt, align=left, sloped}
}

\title{Irregular measurement times in estimating time-varying treatment effects: Categorizing biases and comparing adjustment methods\\\line(1,0){500}}

\author[1]{Wouter M.R. Kant}
\author[2]{Jesse H. Krijthe}
\affil[1]{Institute for Computing and Information Sciences, Radboud University, Nijmegen, The Netherlands}
\affil[2]{Pattern recognition \& Bioinformatics Group, EEMCS, Delft University of Technology, Delft, The Netherlands}

\date{\vspace{-5ex}}

\begin{document}

\maketitle

\begin{abstract}

To estimate the causal effect of treatments that vary over time from observational data, one must adjust for time-varying confounding. A common procedure to address confounding is the use of inverse probability of treatment weighting methods. However, the timing of covariate measurements is often irregular, which may introduce additional confounding bias as well as selection bias into the causal effect estimate. Two reweighting methods have been proposed to adjust for these biases: time-as-confounder and reweighting by measurement time. However, it is currently not well understood in which situations these irregularly timed measurements induce bias, and how the available reweighting methods compare to each other in different situations. In this work, we provide a complete inventarization of all possible backdoor paths through which bias is induced. Based on these paths, we distinguish three categories of confounding bias by measurement time: direct confounding (DC), confounding through measured variables (CMV), and confounding through unmeasured variables (CUV). These categories differ in the assumptions and reweighting methods necessary to adjust for bias and may occur simultaneously with selection bias. Through simulation studies, we illustrate: 1. Reweighting by measurement time may be used to adjust for selection bias and confounding through measured variables; 2. Time-as-confounder may be used to adjust for all categories of confounding bias, but not selection bias; 3. In some cases, the use of a combination of both techniques may be used to adjust for both confounding and selection bias. We finally apply the categorization and reweighting methods on the pre-DIVA\footnote{The preDIVA trial is registered under ISRCTN, registry number ISRCTN29711771. Url: https://www.isrctn.com/ISRCTN29711771.} data set. Adjusting for measurement times is crucial in order to avoid bias, and the categorization of biases and techniques that we introduce may help researchers to choose the appropriate analysis.
\end{abstract}
\author{}

\newpage
\section{Introduction}

In medicine and epidemiology, we are often interested in estimating the causal effect of a treatment that occurs over a prolonged period of time, such as the effect of a lifestyle intervention (e.g. physical activity) on cognitive health \citep{hamer2009physical, kivipelto2013finnish, colcombe2003fitness}. Randomized trials are the standard for estimating such effects. However, setting up a randomized trial to estimate this effect is often difficult or impossible due to large costs, ethical concerns, dropout, and low adherence to treatment. Instead, observational data may be used to draw causal conclusions, provided that we can make certain assumptions.

\begin{figure}[h!]
\centering
\includegraphics[width=0.5\linewidth]{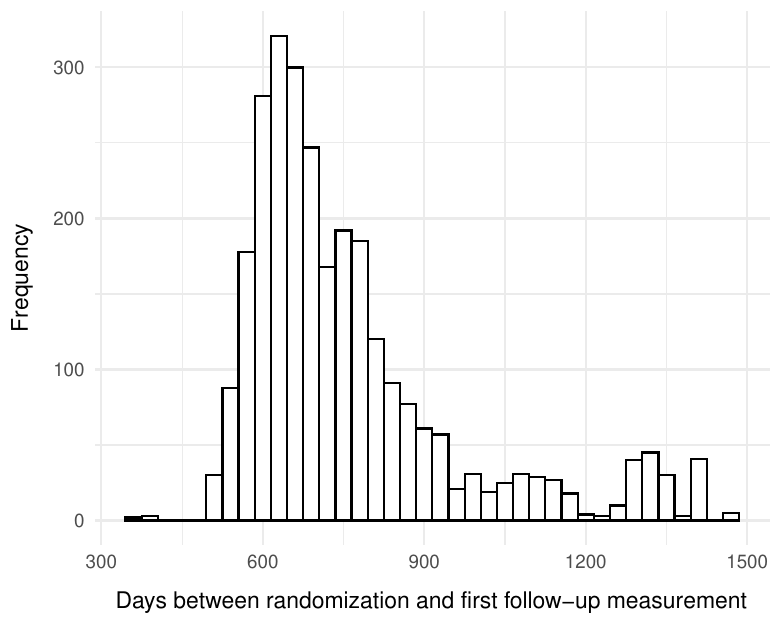}
\caption{\protect\label{fig:MeasurePreDiva}The amount of time between the baseline measurement and the first follow-up measurement in the preDIVA study \citep{ligthart_cardiovascular_2015}.}
\end{figure}

Using observational data to draw causal conclusions comes with its own challenges. A particular challenge is to estimate the effect of a prolonged treatment, due to the potential presence of treatment-affected time-varying confounding (TVC). TVC can occur when treatment decisions influence covariates, which in turn can influence future treatment decisions as well as the outcome of interest \citep{robins1986new}. Adjustment for confounding at baseline in the presence of TVC is not sufficient to estimate the causal effect, and simple regression methods are known to be biased when adjusting for TVC \citep{robins1986new, hernan2009observation}.
Fortunately, several techniques have been developed that are not biased when estimating treatment effects in the presence of TVC, including the so-called g methods. One of the g methods is inverse-probability-of-treatment-weighting (IPTW). IPTW can be applied to estimate the parameters of Marginal Structural Models, which are asymptotically unbiased estimators of the causal effect of a time-varying treatment in the presence of TVC \citep{robins2000marginal, hernan2006estimating}. 
However, when adjusting for TVC using IPTW, unequal measurement times between participants, which we will refer to as irregular measurement times, can cause the causal effect estimate to be biased nonetheless \citep{hernan2009observation, chamapiwa2019application}. There are many reasons why these irregular measurement times may occur, which include practical issues such as capacity of the research team and weekends \citep{pullenayegum_randomized_2022}.   As an example of measurement time irregularity in a typical real-world trial, consider the preDIVA study, which was a lifestyle intervention study in which participants were asked to return in 2-year intervals \citep{ligthart_cardiovascular_2015}. In Figure \ref{fig:MeasurePreDiva}, it can be seen that simply due to logistics, the amount of time between baseline and the first follow-up may vary. Importantly, this was not an issue for estimating the effectiveness of the main intervention. However, if we want to estimate other effects which are not randomized, such as the effect of treatment adherence over time using IPTW, then our estimates may become biased due to these irregular measurement times.

These biases introduced by irregular measurement times can be divided into either selection bias or confounding bias \citep{hernan2009observation}. Selection bias may occur when the analysis is restricted to people who were measured at certain time points. Confounding bias may occur when the time of measurement is dependent on past, potentially unmeasured covariates. In this setting, we will categorize confounding bias into three relevant categories. This categorization will help to understand the source of bias, and to know which methods will be effective in removing this bias. In the past, multiple categorizations of bias have been created for the setting of irregular measurement times. \cite{tan2014regression} introduce a categorization of outcome-observation dependence mechanisms. \cite{gasparini2020mixed} compare methods to adjust for informative visit times that induce bias in longitudinal regression coefficients. \cite{pullenayegum_longitudinal_2016} have categorized visit processes and reviewed methods to adjust for irregular measurement times in longitudinal settings where the outcome is modeled over time. Instead, we construct our categorization of biases in a setting with TVC, where the outcome is modeled at a singular final time point. Using this categorization, we will show that unbiased estimation is possible even in the presence of \textit{patient-driven visits} for the covariate measurements, which may correspond to Pullenayegum and Lim's \textit{Visiting not at random} (VNAR). \cite{goldstein_how_2019} state that an association cannot be induced by informative measurements in a setting without TVC. We show that in the presence of TVC, leaving the measurement time unadjusted for may in fact induce association, even when there is no association between treatment effect and outcome.

We discuss several methods that may be used to adjust for TVC in a setting with irregular measurement times. \cite{hernan2009observation} have introduced a reweighting method in which each individual is reweighted by the probability of being measured at all their respective measurement times, based on their covariate measurements. For ease of reference, in this paper we will call this method \textit{Reweighting by measurement time} (RMT). This method is reminiscent of, but not identical to, \textit{Inverse intensity of visit weighting} (IIVW), developed by \cite{robins1995analysis} and \cite{burv}, which adjusts for outcome-dependent follow-up measurements. \cite{kalia2023estimation} have applied RMT in combination with generalized estimating equations and calibrated weights in order to account for irregular measurement times in addition to individual unmeasured confounders. An alternative method of adjusting for irregular measurements has been proposed by \cite{chamapiwa2019application}, in which the measurement time itself is treated as a confounding variable. We refer to this technique as \textit{time-as-confounder} (TAC).

\begin{figure}[h!]
\centering
\subfloat[][ \label{fig: ConfBiases1} Direct confounding (DC)]
  {\includegraphics[]{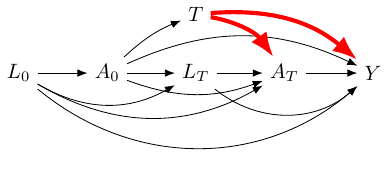}}
    \hfill\\
\subfloat[\label{fig:ConfBiases2} Confounding through measured variables (CMV)]
  {\includegraphics[]{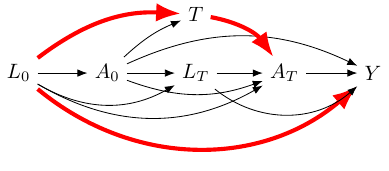}}
\hfill\\
\subfloat[\label{fig:ConfBiases3} Confounding through unmeasured variables (CUV)]
  {\includegraphics[]{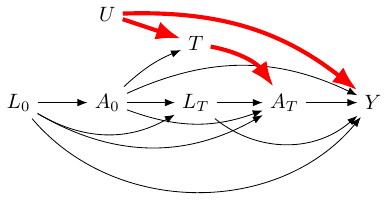}}
\caption{Our categorization of backdoor paths through which bias is introduced into the treatment effect estimate through irregular measurement times, depicted in TAV graphs. Direct Confounding (DC) (a), Confounding through Measured Variables (CMV) (b), and Confounding through Unmeasured Variables (CUV) (c). $L_0$ and $A_0$ denote covariate measurements and treatment decision at baseline, respectively. $T$ denotes the amount of time between the baseline measurement and the follow-up measurement. $L_T$ and $A_T$ denote the covariate measurement at the time of measurement $T$. $Y$ denotes the outcome measurement. The full list of backdoor paths can be found in the supplementary material. The red arrows denote examples of backdoor paths that fall within the categories. \protect\label{fig: AllConfPaths}}
\end{figure}

In this paper, we explore different ways to apply IPTW in settings with irregular measurement times. The paper is structured as follows. Firstly the problem setting is introduced in Section 2. In Section 3 we define a categorization of biases by finding all causal pathways through which confounding bias by irregular measurement times can be introduced. These causal paths are divided into Direct Confounding (DC), Confounding through Measured Variables (CMV) and Confounding Through Unmeasured Variables (CUV), which are visible in Figure \ref{fig: AllConfPaths}. In Section 4 previous methods to adjust for irregular measurement times are introduced, specifically TAC and RMT. We show how these reweighting methods adjust for confounding bias and/or selection bias in different circumstances, and that the use of both techniques simultaneously may be necessary in order to obtain unbiased treatment effect estimates. In Section 5, we describe the simulations that are used to show which techniques are effective in which situations. In Section 6, the real-life preDIVA dataset is used as an example of the benefit of our categorization of biases, and we apply TAC and RMT. To our knowledge, this is the first time that TAC is applied on a real-life dataset. Finally, in Section 7 we discuss the implications of the results presented in this paper, and give recommendations on when to apply which method.

\section{Setting}

\begin{figure}[h!]
    \centering
    \begin{tikzpicture}
        \node (Lt) at (0,0) {$L_{0}$};
        \node (At) at (1.5,0) {$A_{0}$};
        \node (Lt1) at (3,0) {$L_{1}$};
        \node (At1) at (4.5,0) {$A_{1}$};
        \node (Y) at (6,0) {$Y_{T}$};
        \node (U) at (0,1.5) {U};
        
        \path (Lt) edge (At);
        \path (Lt) edge[bend right = 30] (Lt1);
        \path (Lt) edge[bend right = 30] (At1);
        \path (At) edge (Lt1);
        \path (At) edge[bend right  = 20] (At1);
        \path (At) edge[bend left = 30] (Y);
        \path (At1) edge (Y);
        \path (Lt1) edge (At1);
        \path (Lt) edge[bend right = 50] (Y);
        \path (Lt1) edge[bend right = 40] (Y);

        \path (U) edge[] (Lt);
        \path (U) edge[] (Lt1);
        \path (U) edge[bend left = 20] (Y);
        
    \end{tikzpicture}
    \caption{A causal graph where there is treatment-affected time-varying confounding present. $A_0$ and $A_1$ denote treatment decisions at baseline and the first follow-up, respectively. There is an unmeasured variable $U$ that affects covariates $L_0$ and $L_1$ as well as outcome $Y$.}
    \label{fig:simplesetting}
\end{figure}
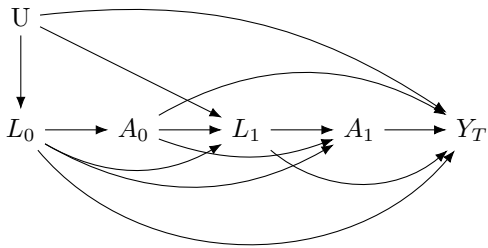

Our goal is to estimate the causal effect of a treatment on an outcome of interest. In our particular case study, our treatment variable is adherence to the treatment protocol. The outcome of interest is whether people had a high or low physical activity afterwards. We intend to estimate the average causal effect of our treatment variable on the outcome variable, also known as the average treatment effect (ATE). Say that there are $K$ measurement points (e.g. doctor’s visits) where covariates are measured. At each of these measurement points, a binary treatment (e.g. low or high adherence) $A_{i,k} \in \{0,1\}$ is determined for each individual $i \in \{1,...,n\}$. Let $T_{i,k} \in \mathbb{R}$ denote the time since baseline at which these measurement points occured. The outcome (e.g. physical activity level) $Y \in \mathbb{R}$ is measured at some time point after the final covariate measurement. The treatment history of an individual during the study is denoted as $\bar{A}_{i,k} = \{A_{i,1}, ..., A_{i,k}\}$. Similarly, any variable with a bar is a denotation of the history of that variable. Let $L_{i,k}$ be the covariates of individual $i$ at measurement $k$, which includes potential confounding variables, such as markers indicating vascular health. Figure \ref{fig:simplesetting} depicts a causal graph that illustrates the causal structure that is assumed for the setting.

The ATE that we want to estimate is defined as follows:

\begin{align}
    \text{ATE} = E[Y^{\bar{a} = \bar{1}}-Y^{\bar{a}_0 = \bar{0}}]
\end{align} in which $Y^{\bar{a}= \bar{1}}$ denotes the (counterfactual) outcomes that would have been observed, had the study population stayed on treatment $1$ for the duration of the study (e.g. stayed physically active for the entire study period). Similarly, $Y^{\bar{a}=\bar{0}}$ denotes the outcomes that would have been observed, had the study population followed treatment $0$ for the duration of the study (e.g. were physically inactive for the entire study period). Ideally, both counterfactual outcomes of each individual would be directly compared to find the treatment effect. However, because each individual cannot experience both possible treatment histories, it is impossible to observe the full counterfactual outcome vectors $Y^{\bar{a}=\bar{1}}$ and $Y^{\bar{a}=\bar{0}}$.

Instead, the `crude' treatment effect may be estimated $ E[Y|\bar{A}=\bar{1}] - E[Y|\bar{A} = \bar{0}]$, using the observed treatment histories and outcomes. However, this will be a biased estimate of the ATE due to confounding by $\bar{L}$. Classic ways to adjust for this confounding, for example by fitting a linear model with $L_0$ and $L_1$ as covariates, will be biased as well. This is because $L_1$ is a mediator of the effect of $A_0$ on $Y$ as well as a common effect of $A_0$ and potential unmeasured variable $U$ \citep{robins1999association}. In order to properly adjust for this confounding, g methods were developed. The g methods include IPTW to estimate marginal structural models, the (parametric) g-formula and g-estimation of structural nested models \citep{hernan_causal_2023}. We restrict our main analysis to IPTW.

\subsection{Irregular measurement times: DT graphs and TAV graphs}

We now introduce the problem of irregular measurement times into this setting. This means that the times at which measurements take place differs between individuals. Rather than a single regime of measurement times that everyone follows, each person has their own individual measurement times. For now, we restrict the analysis to a setting where the number of measurements $K_i$ is the same between individuals.
There are multiple ways in which these irregular measurement times can cause bias \citep{hernan2009observation}. In order to find all possible ways in which they can cause bias, we include the measurement times into the causal graph. We describe two ways to do this: `Discrete time' (DT) graphs and `time-as-variable' (TAV) graphs.

\subsubsection*{Discrete time (DT) graphs}

\begin{figure}[ht]
    \centering
    \includegraphics[width=0.5\textwidth]{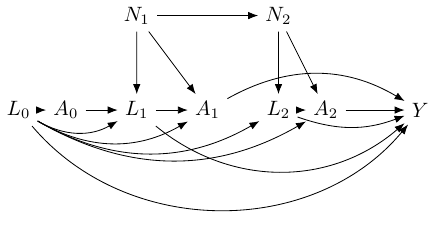}
    \caption{An example of a discrete time (DT) graph, where nodes $N_1$ and $N_2$ indicate whether an individual is measured at that time or not.}
    \label{fig:DTgraph}
\end{figure}

DT graphs are causal graphs where the potential measurement times (which are generally on a continuous scale) are discretized into a finite number of discrete time points. For each of these discrete time points, there is a node for covariates $L$, treatment $A$, as well as an indicator node $N$ which indicates whether an individual was measured at this time or not. In our example, a DT graph where the time of follow-up is discretized into two possible time slots could look as depicted in Figure \ref{fig:DTgraph}. If there is a measurement in the time slot defined for $N_1$, then $N_1=1$, and if there was no measurement within that time slot, then $N_1 = 0$.
In a DT graph, the nodes for treatment ($A_0$, $A_1$ and $A_2$) are easily interpretable as the treatment decision that was made within that discretization of time. A DT graph can also incorporate individuals who were measured a different number of times during the study period. For example, some individuals can have $N_1=1$ and $N_2=1$ (measured at both times 1 and 2), and others could have $N_1=0$ and $N_2=0$ (measured at neither times). For measurement times where $N_{k+1}=0$ (and the individual $i$ is thus not measured), it is often assumed that $A_{i,k+1} = A_{i,k}$ if $N_{k+1} = 0$ \citep{hernan2009observation}. In other words, it is assumed that the treatment that is followed may only change when someone is measured. A downside of this assumption is that the conditional probability of treatment reaches 0 or 1 for parts of the population that have a certain combination of covariate values, because $P(A_2=a_2|A_1=a_1,A_0 = a_0,N_2=0) = 0$ for any $a_2 \neq a_1$. We will observe in Section 4 that this phenomenon violates a necessary assumption (namely, the positivity assumption) for some reweighting methods.

\subsubsection*{Time-as-variable (TAV) graphs}

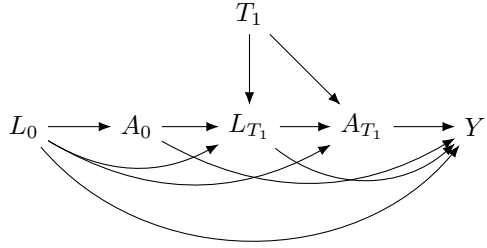
\begin{figure}[h!]
\centering
\begin{tikzpicture}
    \node (Lt) at (0,0) {$L_0$};
    \node (At) at (1.5,0) {$A_0$};
    \node (Ot1) at (3,1.5) {$T_1$};
    \node (Lt1) at (3,0) {$L_{T_1}$};
    \node (At1) at (4.5,0) {$A_{T_1}$};
    \node (Y) at (6,0) {$Y$};

    \path (At) edge[bend right = 30] (Y);
    \path (At1) edge (Y);
    \path (Lt) edge (At);
    \path (Lt) edge[bend right = 30] (Lt1);
    \path (Lt) edge[bend right = 30] (At1);
    \path (At) edge (Lt1);
    \path (Lt1) edge (At1);
    \path (Lt) edge[bend right = 50] (Y);
    \path (Lt1) edge[bend right = 40] (Y);

    \path (Ot1) edge (Lt1);
    \path (Ot1) edge (At1);
\end{tikzpicture}.
\caption{A TAV graph with treatment-affected time-varying confounding present, where $T_{1}$ represents the time at which follow-up 1 took place.}
\label{fig:TAV graph}
\end{figure}

In a TAV graph, as previously used by \cite{chamapiwa2019application}, the time of measurement itself is depicted as a node into the graph. This is visible in Figure \ref{fig:TAV graph} Examples of TAV graphs can be see in Figure \ref{fig: AllConfPaths}. $T_1$ denotes the time at which the follow-up took place, $A_{T_1}$ denotes the treatment decision that was made at time $T_1$, and $L_{T_1}$ denotes the covariates that were measured at time $T_1$.

There are two benefits to depicting the graph in this way. First of all, it is no longer necessary to make the assumption that $A_{i,k+1} = A_{i,k}$ if $N_{k+1} = 0$, because $N$ is no longer a part of the graph. Because of this, the positivity assumption violation that was mentioned earlier is no longer an issue. The second benefit is that the time does not have to be discretized, because $T_1$ can be a continuous variable. However, a limitation of the TAV graph is that the interpretation of treatment decision $A_{T_1}$ in the TAV graph is different from treatment decisions $A_1$ and $A_2$ in the DT graph. $A_{T_{1i}}$ denotes the treatment decision at the individuals' time of measurement  $T_{1i}$, rather than the treatment decision at a specific point in time.

In the following sections, DT and TAV graphs will be used to show different categorizations of confounding and selection bias, and we will connect this to various techniques to adjust for bias, that assume either a DT or TAV graph.

\newpage

\section{Confounding bias and selection bias}

Now that irregular measurement times are incorporated into the graph, the next step is to uncover when bias is caused by irregular measurement times. This happens when there is a causal pathway that introduces spurious correlation between treatment history $\bar{A}$ and outcome $Y$, which is referred to as a backdoor path \citep{neuberg2003causality}. Obvious examples of backdoor paths are caused by time-varying confounder history $\bar{L}$, and may be successfully adjusted for using IPTW. However, there are still many situations in which the variation in measurement time opens up backdoor paths that are not there in situations with regular measurement times. This means that our causal effect estimation of the ATE will be biased, even after applying IPTW. These paths may be divided into confounding bias, which are the result of a common factor influencing both treatment and outcome, and selection bias, which is the result of conditioning on a variable.

\subsection{Categorization of confounding bias}
\begin{figure}[h!]
\centering
\subfloat[width=0.45\linewidth][ \label{fig: Procedure1} Step 1: Start with an example graph. We have removed paths that may either be adjusted for using IPTW, or violate the exchangeability assumption even under euqal measurement times.]
  {\includegraphics[]{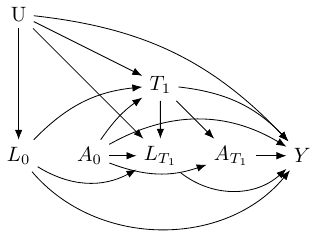}}
\hspace{2em}
\subfloat[width=0.45\linewidth][\label{fig: Procedure2} Step 2: Find all possible backdoor paths. Firstly a root node is selected (this variable is encircled, in this case U). The backdoor path consists of a path towards treatment $A$ (thick arrow, dashed), and a path towards outcome $Y$ (thick arrow, not dashed). One backdoor path is shown.]
  {\includegraphics[]{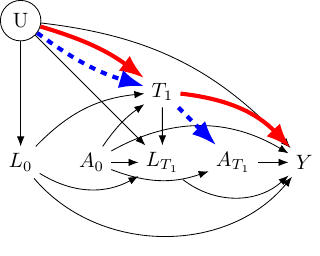}}
\hfill\\
\subfloat[width=0.45\linewidth][\label{fig: Procedure3} Step 3: Remove backdoor paths of which two downstream paths flowed through the same variable. The backdoor path shown in Step 2 is removed in favor of the backdoor path shown above, where $T_1$ is the new root node.]
  {\includegraphics[]{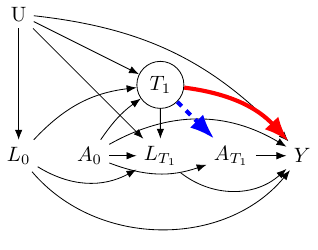}}
\caption{Our 3-step procedure to find all the relevant backdoor paths caused by confounding bias. In the image a TAV graph is shown. The same procedure was followed for DT graphs.}
\end{figure}

We introduce an overview and categorization of backdoor paths due to confounding bias due to irregular measurement times. The categorization that we find is valid for both TAV and DT graphs. The following procedure is used to create the categorization. The procedure has three steps:
\begin{enumerate}
    \item Constructing the starting graph (see Figure \ref{fig: Procedure1})
    \item Finding all backdoor paths between $\bar{A}$ and $Y$ (See Figure \ref{fig: Procedure2})
    \item Removing redundant components of the backdoor paths (See Figure \ref{fig: Procedure3})
\end{enumerate} 

See the Appendix for more details on the procedure used to find all of the backdoor paths, and see the supplementary material for a list of all backdoor paths and which categorization of bias they belong to.

\subsubsection*{Categorization of biases}

The relevant backdoor paths can be divided into three categories. Let $A \rightarrow B$ denote that there is a direct downstream path from variable $A$ to $B$, that does not flow through another variable acting as a mediator. Let $A \bar{\rightarrow} B$ denote the entire downstream path from variable $A$ to variable $B$. Let $R$ be the root node of a backdoor path. Let $M$ be any measured variable, and let $U$ be any unmeasured variable. By definition, all backdoor paths have the requirement that $R \bar{\rightarrow} A$ and $R \bar{\rightarrow} Y$ exist. Then, the three categories of confounding bias are defined as follows for TAV graphs:

\begin{itemize}
    \item `Direct confounding' (DC): $R=T$, $R\rightarrow A$
    \item `Confounding through measured variables' (CMV): $M \rightarrow T \in R\bar{\rightarrow} A$
    \item `Confounding through unmeasured variables' (CUV): $R=U$, $U\rightarrow T \in R \bar{\rightarrow} A$
\end{itemize}

Similarly, the three categories of bias are defined as follows for DT graphs:

\begin{itemize}
    \item `Direct confounding' (DC): $R=N$, $R \rightarrow A$
    \item `Confounding through measured variables' (CMV): $M \rightarrow N \in R\bar{\rightarrow} A$
    \item `Confounding through unmeasured variables' (CUV): $R=U$, $U \rightarrow N \in R \bar{\rightarrow} A$
\end{itemize}

Examples of a path in each category can be found in Figure \ref{fig: AllConfPaths} for TAV graphs. The categorizations are useful in deciding how to adjust for the confounding bias caused by irregular measurements,
because, as we will see, different estimation methods are effective in adjusting for different categories of backdoor paths. An additional result of interest is that in TAV graphs, all three categories of bias are only present if measurement time $T$ affects treatment decision $A$. Similarly, in DT graphs the categories of confounding bias are only present if indicator variable $N$ affects treatment decision $A$. 

\subsection{Selection bias}

\begin{figure}[ht]
    \centering
    \includegraphics[]{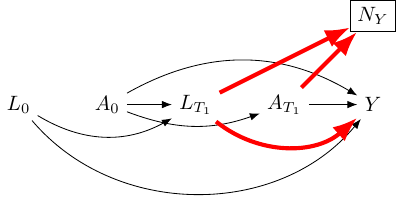}
    \caption{A causal graph where there is selection $N_Y$ on the outcome being measured, which induces selection bias by conditioning on a common effect of a confounding variable and treatment.}
    \label{fig:selectionbias}
\end{figure}

Selection bias due to irregular measurement times may occur when certain individuals are excluded from the analysis based on the time of their observations. We will consider a single example of selection bias to show that confounding bias and selection bias may occur simultaneously. Say that we are interested in estimating the outcome at a specific point in time. Let $N_{Y}$ be an indicator of whether the outcome was measured at that time. Figure \ref{fig:selectionbias} shows how $N_Y$ may induce a backdoor path. By only including persons with $N_Y = 1$, the analysis is conditioned on $N_Y = 1$. In this case, both treatment and confounder levels affect $N_Y$, which means that $N_Y$ acts as a collider. By conditioning on a collider we may open up backdoor paths, introducing bias into the treatment effect estimate \citep{hernan2009observation}.

\section{Adjustment methods}

There are several ways to modify the application of IPTW in order to adjust for either confounding bias and selection bias under DC, CMV and CUV. First we will briefly introduce IPTW. Then, we will highlight the different methods and study which categories of confounding bias they can adjust for.

\subsection{IPTW}

When using IPTW, the inverse-probability-of-treatment-weights are used to reweight the observations to create a `pseudo-population' in which there is no effect of confounding factors $\bar{L}$ on treatment decisions $\bar{A}$ \citep{robins2000marginal}. An individual's stabilized weight is:

\begin{align}
    \text{SW}_i^{\text{IPTW}} = \prod_{k=0}^K \frac{P(A_{i,k}|\bar{A}_{i,k-1}) }{P(A_{i,k}|\bar{A}_{i,k-1},\bar{L}_{i,k})}.
\end{align}

In the pseudopopulation, the ATE $E[Y^{\bar{a} = \bar{1}} - Y^{\bar{a} = \bar{0}}]$ may be consistently estimated by $E[Y|\bar{A} = \bar{1}] - E[Y|\bar{A} = \bar{0}]$. However, this is subject to the following causal assumptions of sequential conditional exchangeability, positivity and consistency. These are, respectively:

\begin{align}
    &Y^{\bar{A}}  \perp\!\!\!\perp  A_k | \bar{A}_{k-1}, \bar{L}_k\\
    &P(A_k|\bar{A}_{k-1},\bar{L}_k) > 0 \text{  for all  } A_k, \bar{A}_{k-1}, \bar{L}_k \text{ where } P(\bar{A}_{k-1}, \bar{L}_k) > 0\\
    &Y^{\bar{a}} = Y \text{  if  } A = \bar{a}
\end{align}
Additionally, it must be assumed that there is no model misspecification of $P(A_{i,k}|\bar{A}_{i,k-1})$ and $P(A_{i,k}|\bar{A}_{i,k-1},\bar{L}_{i,k})$.

$\text{SW}^{\text{IPTW}}$ effectively removes the causal paths from time-varying confounders $L$ to the treatment decision $A$. However, in the case of DC and CUV, this does not block all backdoor paths between treatment decisions $A$ and outcome $Y$. This can be seen by examining Figure \ref{fig: ConfBiases1} and \ref{fig:ConfBiases2}. Therefore, the exchangeability assumption is violated, and the ATE estimate will be biased.

CMV, visible in Figure \ref{fig:ConfBiases3}, does not cause bias when applying $\text{SW}^{\text{IPTW}}$, if the variables $L$ causing CMV are the same variables $L'$ that are in the denominator of $\text{SW}^{\text{IPTW}}$. This is because applying $\text{SW}^{\text{IPTW}}$ effectively removes the causal paths from $L$ to $A$. $T$ is a mediator on this path and may in this case be ignored. 

However, this backdoor path may be accidentally opened up again if caution is not taken. For example, when a researcher notices a difference in measurement times, they may consider to add the variable $T$ into Equation (2), because the probability of treatment could be estimated more precisely by including the variable $T$ into the conditioning set of both the numerator and denominator. This would result in the following weights calculation for a TAV graph:

\begin{align}
    \text{SW}_i^{\text{IPTW-T}} = \prod_{k=0}^K \frac{P(A_{k,i}|\bar{A}_{{k-1},i}, \bar{T_k}) }{P(A_{k,i}|\bar{A}_{{k-1},i},\bar{T_k},\bar{L}_{k,i})}.
\end{align}
By including $\bar{T}$ into both numerator and denominator of the calculation of stabilized weights, the possible confounding caused by that variable is no longer adjusted for. This opens up the path $L_0 \rightarrow T_1 \rightarrow A_{T_1}$. As a consequence, the backdoor path causing CMV is opened up, and there may be bias under CMV when applying $\text{SW}_i^{\text{IPTW-T}}$.

\subsection{Reweighting by measurement time}
\cite{hernan2009observation} introduced stabilized weights to adjust for a dynamic observation plan $\text{SW}^{N}$ when using DT graphs. Let $T_{\text{max}}$ denote the latest possible time for a measurement to have taken place. The weights for the probability of treatment are calculated by
\begin{align}
    \text{SW}_i^{A(N)}= \prod_{t=0}^{T_{\text{max}}} \frac{P(A_{i,t}|\bar{N}_{i,t},\bar{A}_{i,t-1}) }{P(A_{i,t}|\bar{N}_{i,t},\bar{A}_{i,t-1},\bar{L}_{i,t})}.
\end{align}
These weights adjust for the causal paths of covariates $\bar{L}$ into treatment decisions $\bar{A}$. Additionally, the weights for the probability of being observed at any given time point are calculated by
\begin{align}
    \text{SW}_i^N = \prod_{t=0}^{T_{\text{max}}}\frac{P(N_{i,t+1}|\bar{N}_{i,t},\bar{A}_{i,t}) }{P(N_{i,t+1}|\bar{N}_{i,t},\bar{A}_{i,t},\bar{L}_{i,t})}.
\end{align}
These weights adjust for the causal paths from covariates $\bar{L}$ into the measurement indicators $N$.
Then, the final weights are
\begin{align}
    \text{SW}_i^{A,N} = \text{SW}_i^{A(N)} \cdot \text{SW}_i^N,
\end{align}
where for every time point where $N=0$, $\text{SW}^{N}= 1$. Combining the weights in this way creates a pseudopopulation in which both the causal edges from $\bar{L}$ into $\bar{A}$ and the edges from $\bar{L}$ into $\bar{N}$ are removed.

This method is only defined for DT graphs. However, it is possible to also use this method in TAV graphs. The weights would look as follows:
\begin{align}
    \text{SW}_i^T = \prod_{k=0}^{K}\frac{P(T_{i,k}|\bar{T}_{i,k-1},\bar{A}_{\bar{T},i,k-1}) }{P(T_{i,k}|\bar{T}_{i,k-1},\bar{A}_{\bar{T},i,k-1},\bar{L}_{\bar{T},i,k-1})}.
\end{align}

Rather than reweighting at each possible time point $1... T$, each individual is only reweighted at each measurement time until $1... K$, where $K$ is the total number of measurements. For ease of reference, we refer to the use of this technique as `Reweighting by measurement time' (RMT).

RMT makes the following assumptions about $T$ \citep{hernan2009observation}:

\begin{enumerate}[label=($O_{{\arabic*}}$)]
    \item
    No unmeasured variables affecting $T$ and $Y$
    \item
    No direct effect of $T$ on either $L$ or $Y$
    \item The models for $P(T|\bar{T},\bar{A})$ and $P(T|\bar{T},\bar{A},\bar{L})$ are correctly specified
\end{enumerate}

When using IPTW and RMT, the stabilized weights calculation would look as follows:

\begin{align}
    \text{SW}_i^{\text{RMT}} = \text{SW}_i^{\text{IPTW-T}} \cdot \text{SW}_i^T.
\end{align}

RMT effectively eliminates causal edges towards T from the measured variables in the dataset in the TAV graph. Some, but not all, backdoor paths causing confounding bias may be eliminated by this adjustment. As is visible in figure \ref{fig: AllConfPaths}, CMV is the only category of confounding bias that depends on the causal edge from a measured variable towards $T$. As a consequence, RMT may only be used to adjust for CMV. CMV is also the only category of confounding bias that fulfills both assumptions $(O_1)$ and $(O_2)$, because DC breaks assumption $(O_2)$, and CUV breaks assumption $(O_1)$.

\subsection{Time-as-confounder}

Another way to adjust for confounding bias due to measurement time is to adjust for the measurement time as if it were a time-varying confounder, as introduced by \cite{chamapiwa2019application}, using a TAV graph. We refer to this as Time-as-confounder (TAC). This method allows for estimation of the average treatment effect in a population with the same distribution of treatment times as your data \citep{hernan2009observation}. When using IPTW, using TAC would look as follows when using a TAV graph:

\begin{align}
    \text{SW}_i^{\text{TAC}} = \prod_{k=0}^K \frac{P(A_{T_k,i,k}|\bar{A}_{\bar{T},i,k-1}) }{P(A_{T_k,i,k}|\bar{A}_{\bar{T},i,k-1},\bar{L}_{\bar{T},i,k}, \bar{T}_{i,k})}.
\end{align}

This method adjusts for the causal path $\bar{T}\rightarrow \bar{A}_{\bar{T}}$. As is visible in Figure \ref{fig: AllConfPaths}, all three categories of confounding bias DC, DUV and DMV are dependent on this path. Therefore, unbiased estimation of the ATE is possible in all three categories of confounding bias.

When the population is reweighted using $\text{SW}^{{\text{TAC}}}$, the causal paths from $\bar{T}$ into $\bar{A}_{\bar{T}}$ are adjusted for. An important consideration is therefore that $\bar{A}_{\bar{T}}$ is defined as the history of treatment decisions at the moments of being measured. This means that we may only unbiasedly estimate the ATE of treatment regimes expressed in terms of $\bar{A}_{\bar{T}}$. However, these treatment regimes can be hard to interpret, because for individuals with differing values of measurement times $T_1$ and $T_2$, the amount of time that they have followed the treatment may differ, even if the treatment decisions were the same. Therefore, it might be of interest to translate the hard-to-interpet $\bar{A}_T$, defined as the treatment decisions at the times of being measured, into the easily interpretable $\bar{A}$, defined as the treatment that was followed at discretized points in time.

For example, say that we are interested in estimating the ATE of following treatment regime $A=1$ for a certain amount of time during the study, summarized as $\sum{\bar{A}}$. We can express this sum in terms of $A_T$ as follows:

\begin{align}
    \sum{\bar{A}} = \sum_{k=1}^{K_i} (T_k-T_{k-1})\cdot A_{T_{k-1}}.
\end{align}

This shows that the measurement times history $\bar{T}$ is necessary in order to translate $\bar{A}_T$ into $\bar{A}$. What this means is that just like the treatment decision $A_T$, the measurement time $T$ must be unconfounded after reweighting the population. The consequence of this is that the estimate of the ATE will be biased under CUV, where it is unavoidable that $T$ is confounded.

However, there are exceptions where the measurement time history $\bar{T}$ is not necessary in order to translate $\bar{A}_T$ into $\bar{A}$. This is the case if, for example, we want to estimate the ATE of having stayed on a certain treatment regime for the entire study period, where every treatment decision $A_{T_{k-1}} = 1$. Then, $\sum_{k=1}^{K_i} (T_j-T_{k-1})\cdot A_{T_{k-1}} = \sum{\bar{A}} = T_Y-T_0$. Thus, the interpretation of the total amount of treatment taken is independent of the measurement times. The same argument goes for any treatment regime where the $\bar{A}_T$ may be translated into $\bar{A}$ without the use of $\bar{T}$. We call these treatment histories `Measurement time-agnostic'.

\subsection{Combining TAC and RMT}

\begin{figure}
    \centering
    \begin{tikzpicture}
        \node (Lt) at (0,0) {$L_0$};
        \node (At) at (1,0) {$A_0$};
        \node (Ot1) at (2.5,1.5) {$T_1$};
        \node (U) at (0,3) {U};
        \node (Lt1) at (2.5,0) {$L_{T_1}$};
        \node (At1) at (4,0) {$A_{T_1}$};
        \node (Y) at (5.5,0) {$Y$};
    
        \path (At) edge[bend left = 30] (Y);
        \path (At1) edge (Y);
        \path (Lt) edge (At);
        \path (Lt) edge[bend right = 30] (Lt1);
        \path (Lt) edge[bend right = 30] (At1);
        \path (At) edge (Lt1);
        \path (At) edge[bend right  = 20] (At1);
        \path (Lt1) edge (At1);
        \path (Lt) edge[bend right = 50] (Y);
        \path (Lt1) edge[bend right = 40] (Y);
    
        \path (Ot1) edge[red, bend left = 22] (At1);
        
        \path (At) edge[bend left = 10]  (Ot1);
        \path (U) edge[red] (Ot1);
        \path (U) edge[red, bend left = 20] (Y);

        \node (Ot2) at (5.5, 1.5) [rectangle, draw] {$T_2$};
        \path (At1) edge[red] (Ot2);
        \path (Lt1) edge[red] (Ot2);
    \end{tikzpicture}
    \caption{A causal graph wherein the use of $\text{SW}^{\text{TAC+RMT}}$ is necessary to remove both the selection bias as well as the confounding bias.}
    \label{fig: Conf+Select}
\end{figure}
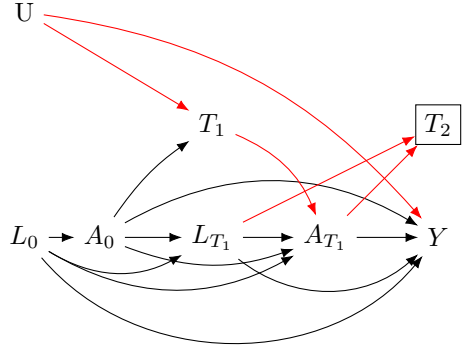

TAC has the advantage that it may adjust for DC and CUV. In contrast, RMT has the advantage that it may adjust for selection bias. However, there are settings in which there is both DC, CUV and selection bias present, where it might be advantageous to apply both techniques. For example, if we are interested in the outcome at a specific time point, then we are effectively conditioning our analysis on that the final measurement time, $T_{K+1}$, has a certain value. A causal graph of this scenario, where CUV is present simultaneously with selection bias, is shown in Figure \ref{fig: Conf+Select}. 
Then, the weights calculation would look as follows:

\begin{align}
    \text{SW}_i^{\text{TAC+RMT}} = &\prod_{k=0}^K \frac{P(A_{T_k,i,k}|\bar{A}_{\bar{T},i,k-1}) }{P(A_{T_k,i,k} |\bar{A}_{\bar{T},i,k-1},\bar{L}_{\bar{T},i,k}, \bar{T}_{i,k})} \\
    & \cdot \frac{P(T_{K+1}|\bar{T}_{i,K},\bar{A}_{\bar{T},i,K}) }{P(T_{K+1}|\bar{T}_{i,K},\bar{A}_{\bar{T},i,K},\bar{L}_{\bar{T},i,K})}. \notag
\end{align}

This may only be done in settings corresponding to a TAV graph, because the use of $\text{SW}^{{\text{TAC}}}$ is restricted to TAV graphs. 

\section{Simulations}

To examine the bias and variance that result from using the different reweighting methods in the different categories of irregular measurement time confounding and selection bias that we identified, we set up a number of simulation scenarios.

Datasets were generated according to TAV causal graphs. All variables were generated conditional on preceding variables within the causal graph. $Y$ is a continuous outcome variable generated according to a normal distribution $Y \sim N(\mu = \beta_Y + \beta_{A_1Y}A_1 + \beta_{A_{1}Y}A_{1} + \beta_{L_0Y}L_0 + \beta_{L_{1}Y}L_{1} + \beta_{UY}U, \sigma^2 = 0.5)$. All other variables, including the follow-up measurement time $T_1$, are binary variables. This means that there are two possible measurement times: `early' ($T_1 = 0$) and `late' ($T_1 = 1$). These variables were generated according to bernoulli distributions, where the probability parameter was generated according to an inverse logit transformation of a linear model. For example, the generative distribution of $L_{1}\sim \text{Bernoulli}(\text{logit}^{-1}(\beta_{L_{1}} + \beta_{UL_{1}}U + \beta_{L_kL_{1}}L_0 +\beta_{A_0L_{1}}A_0 ))$.  For a complete specification of the parameter settings, see the supplementary document. In order to simulate the backdoor paths within DC, CMV and CUV, we use different sets of slopes and intercepts for the parameters. For example, to simulate a causal path from $A_k$ to $L_{k+1}$ within the causal graph, we simply set the slope $\beta_{A_kL_{k+1}}$ to a non-zero value. The graphs used to generate data can be found in the supplementary material.

The simulated populations were reweighted using several different sets of stabilized weights. In Scenarios 1, 2 and 3, which simulate DC, CMV and CUV, respectively, the reweighting methods used were $\text{SW}^{\text{IPTW}}$, $\text{SW}^{{\text{IPTW-T}}}$, $\text{SW}^{{\text{TAC}}}$, $\text{SW}^{{\text{RMT}}}$. For Scenario 4, which simulates CUV and selection bias, we included $\text{SW}^{{\text{TAC+RMT}}}$. For each set of stabilized weights, and for each scenario, we attempt to estimate the ATE by $\hat{\text{ATE}} = \hat{E}[Y|\bar{A}=1] - \hat{E}[Y|\bar{A} = 0]$ after reweighting the population. These are compared to the `true' ATE, which is calculated by generating both counterfactual time-series explicitly (whilst resampling the random error term for $Y$), and calculating the average difference. Bias was defined as as the average difference between the estimated $\hat{\text{ATE}}$ and the true ATE. Variance was defined as the variance of $\hat{\text{ATE}}$. 10000 datasets were simulated containing 2000 simulated individuals.

\subsubsection*{Scenario 1: Direct confounding (DC)}

In the first scenario, we examine the performance of the reweighting methods under DC. The DC scenario that was chosen is one where the timing of the follow-up measurement has a direct causal effect on the treatment decision. Additionally, the timing of the follow-up measurement has a direct positive effect on the outcome, as well as a negative effect on time-varying confounder $L_{T_1}$.
As can be seen in Table \ref{tab:Table}, $\text{SW}^{\text{IPTW}}$, $\text{SW}^{{\text{IPTW-T}}}$, $\text{SW}^{{\text{RMT}}}$ and are clearly biased, whereas $\text{SW}^{{\text{TAC}}}$ was unbiased. This coincides with what we expected to see. In this scenario, $\text{SW}^{{\text{TAC}}}$ has the lowest standard deviation of the reweighting methods.

\subsubsection*{Scenario 2: Confounding through measured variables (CMV)}

In the second scenario we compared the performance of the reweighting methods under confounding through unmeasured variables (CMV). Here, a scenario was chosen where an unmeasured variable impacts both $L_0$ and $Y$. $L_0$ turn affects the measurement time, which in turn affects treatment decision $A_1$, and $U$ affects $Y$ directly. In addition, $L_0$ affects $Y$ directly and through mediator $L_{T_1}$.
It can be observed in Table \ref{tab:Table} that in this scenario, it is only $\text{SW}^{{\text{IPTW-T}}}$ that results in a biased estimation of the ATE, whereas the other reweighting methods are unbiased. It may be observed here that $\text{SW}^{{\text{TAC}}}$ has a slightly higher standard deviation than the other reweighting methods, which may be explained by positivity issues due to the large causal effect of $T$ on $A_1$.

\subsubsection*{Scenario 3: Confounding through unmeasured variables (CUV)}

For the third scenario, we simulated confounding through unmeasured variables (CUV). Here, the unmeasured variable $U$ directly impacts measurement time, which affects treatment decision, and $U$ additionally affects the outcome $Y$ directly. $U$ additionally affects $Y$ through variables $L_0$ and $L_{T_1}$.
We observed that $\text{SW}^{{\text{TAC}}}$ clearly has the lowest percentage bias out of the discussed reweighting methods, as is visible in Table \ref{tab:Table}. However, the variances of $\text{SW}^{{\text{TAC}}}$ is again slightly higher, which may again explained by the large effect of measurement time on treatment decision.

\subsubsection*{Scenario 4: CUV + Selection bias}

\begin{figure}[ht]
\centering
\includegraphics[width=0.5\linewidth]{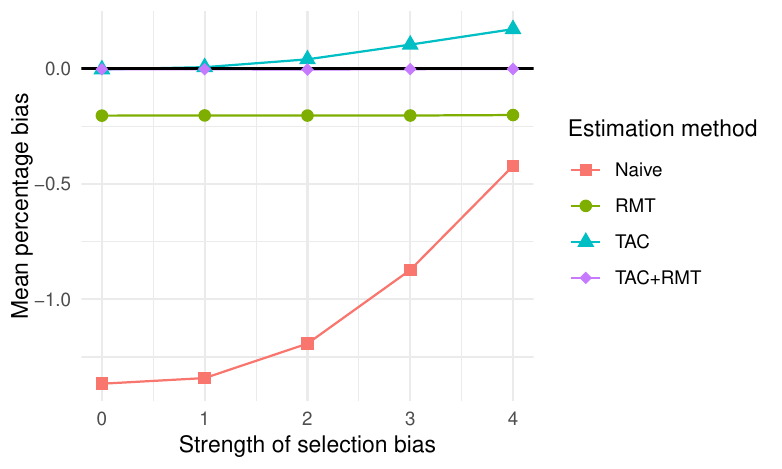}
\caption{\protect\label{fig:ConfoundSelect}  A graph showing the results of Scenario 4, where Confounding Through Unmeasured Variables (CUV) is simulated simultaneously with selection bias. The strength of selection bias by parameter $\gamma$ as described in Section 5 is plotted over the horizontal axis.}
\end{figure}

Selection bias was introduced into the experiments by simulating data according to the same causal graph and parameter settings as Scenario 3 (CUV), while adding an additional variable denoting the time of the outcome variable $T_2$, which indicates whether the outcome of an individual is measured or not. This scenario is depicted in Figure \ref{fig: Conf+Select}. The variable $T_2$ was generated according to $T_2 \sim \text{logit}^{-1}(2-\gamma \cdot A_{T} + \gamma \cdot L_T)$. $\gamma$ was set to several different values as shown. In Figure \ref{fig:ConfoundSelect} we show that when applying $\text{SW}^{{\text{TAC+RMT}}}$ the percentage bias remains lowest for different levels of selection bias.

\begin{table}
\small \sf \centering
\caption{\protect\label{tab:Table}The model performance results of the simulations described in Section 5. Scenario 1 simulates Direct Confounding (DC), Scenario 2 simulates Confounding through measured variables (CMV), Scenario 3 simulates Confounding through unmeasured variables (CUV). 10000 replications have been done, with a sample size of 2000. See the supplementary material for the exact parameter settings.}
\begin{tabular}{ c l c c c c c }
\toprule
\textbf{}                         & Technique & Estimate & True value  & \% bias & SD   & MSE         \\
\midrule
Scenario 1 (DC)  & Naive estimate       & 2.2770 & 2.7903 & -18.3962 & 0.0725  & 0.2687 \\
              & $\text{SW}^{{\text{IPTW}}}$      & 3.1333 & 2.7903 & 12.2949 & 0.0869  & 0.1252  \\
              & $\text{SW}^{{\text{IPTW-T}}}$    & 2.9681 & 2.7903 & 6.3740 & 0.0708 & 0.0366   \\
              & $\text{SW}^{{\text{TAC}}}$       & 2.7896 & 2.7903 & -0.0259 & 0.0678  & 0.0046  \\
              & $\text{SW}^{{\text{RMT}}}$       & 2.8912 & 2.7903 & 3.6171 & 0.0836 & 0.0172   \\
              
\midrule
Scenario 2 (CMV)               & Naive estimate       & 0.9698 & 2.1749 & -55.4107 & 0.0735  & 1.4578 \\
               & $\text{SW}^{{\text{IPTW}}}$      & 2.1807  & 2.1749    & 0.2629 & 0.0467 & 0.0022    \\
               & $\text{SW}^{{\text{IPTW-T}}}$    & 1.8572    & 2.1749      & -14.6068  & 0.0533 & 0.1038 \\
               & $\text{SW}^{{\text{TAC}}}$       & 2.1714     & 2.1749    & -0.1625 & 0.0675 & 0.0046  \\
               & $\text{SW}^{{\text{RMT}}}$        & 2.1735    & 2.1749     & -0.0648 & 0.0519 & 0.0027 \\
\midrule

Scenario 3 (CUV) & Naive estimate       & 0.5795 & 2.0226 & -71.3463 & 0.1041  & 2.0932 \\
 & $\text{SW}^{{\text{IPTW}}}$      & 2.1439 & 2.0226     & 5.998 & 0.0843 & 0.0218 \\
 & $\text{SW}^{{\text{IPTW-T}}}$    & 1.7225     & 2.0226     & -14.8358 & 0.0777 & 0.0961\\
 & $\text{SW}^{{\text{TAC}}}$       & 2.0177    & 2.0226     & -0.2430 & 0.1085 & 0.0118\\
 & $\text{SW}^{{\text{RMT}}}$       & 2.0178     & 2.0226     & -2.2380 & 0.0926 & 0.0086\\
\bottomrule
\end{tabular}
\end{table}

\section{Real-world application}

To illustrate the usefulness of the categorization and reweighting methods, we will consider an analysis of the effectiveness of an intervention on physical activity in the real-world preDIVA trial. The preDIVA trial is a cluster-randomized trial with the primary aim of finding a causal effect of lifestyle changes on cardiovascular health, dementia and disability \citep{ligthart_cardiovascular_2015}. The study population consisted of people aged 70 to 78 years. Over a period of 6 years, people in the treatment arm were assigned 4-monthly visits to a practice nurse. At these visits, participants received advice on healthy lifestyle based on national guidelines. However, not all participants in the treatment arm visited all of these 4-monthly visits. In order to assess the effectiveness of these visits, and as an example of an application of the discussed techniques, we consider the causal estimand of interest to be the effect of `treatment adherence' (defined as having attended at least 66\% of the assigned nurse visits) on the lifestyle of the participants, expressed as whether they fulfilled the World Health Organization norm for physical activity. 1070 participants in the treatment group had their physical activity and adherence recorded at every follow-up measurement within the timeframe of the analysis.

Follow-up measurements were scheduled every 2 years. However, as seen in figure \ref{fig:MeasurePreDiva}, the actual timing of these follow-ups differs per individual. Let $T_1$ be the amount of days between randomization and the first follow-up. Due to adherence being measured at the time of follow-up, our definition of treatment is automatically defined as the treatment at follow-up time $A_T$, which naturally fits more with a TAV graph, rather than a DT graph. Let $A_{T_1}$ denote whether the participant has attended at least 66\% of the assigned nurse visits between randomization and the first follow-up, and let $A_{T_2}$ denote the same for the period between the first and second follow-up. We want to know whether participants who stayed adherent ($A_{T_1}=1$ and $A_{T_2} = 1$), also had a higher physical activity at the time of second follow-up than people who were not adherent for the entire period ($A_{T_1} = 0$ and $A_{T_2} = 0$). For ease of notation, we shorten these to $\bar{A} = \bar{1}$ and $\bar{A} = \bar{0}$, respectively. Our causal estimand is therefore
\begin{align}
    E[Y^{\bar{A} = \bar{1}} - Y^{\bar{A}=\bar{0}}]
\end{align} in which $Y$ is a binary variable denoting whether an individual fulfills the World Health Organization norm for physical activity at the second follow-up.

Adherence to treatment is not randomized and is likely to be confounded. Furthermore, because it is a treatment occurring over a prolonged period of time, the potential of time-varying confounding should be considered. In order to adjust for potential time-varying confounding, we may use IPTW. The baseline variables that were adjusted for are age, gender and general practicioner. Additionally the potential time-varying confounder of physical activity at baseline and the first follow-up was adjusted for. It is unlikely that all of the time-varying confounding was adjusted for, because levels of physical activity as well as adherence may change in between the measurements of the study. Possible violations of the positivity assumption were checked by examining the distribution of IPTW weights. Violations of the consistency assumption are possible due to the aforementioned definition of adherence, which means that there are different levels of adherence within the groups denoted as `adherent' and `non-adherent'.

Different categories of confounding bias may be present due to $T_1$. Identifying which categories are present is important, because this may impact which technique is applicable. Direct confounding (DC) might occur if the occurrence of the follow-up meeting itself has a motivating impact on the participants. Confounding through measured variables (CMV) might occur if a measured factor, such as which general practitioner someone was treated by, impacts both measurement time and outcome. Confounding through unmeasured variables (CUV) might occur if an unobserved latent variable, such as ‘motivation’, could impact both timing of measurement and outcome. Thus, none of the categories of confounding bias can be excluded. Therefore, TAC is likely to be the safest estimation method.

The naive estimate without reweighting was compared to IPTW, TAC and RMT. For each of the reweighing methods, inverse probability of censoring weights (IPCW) were calculated using the same set of variables as for IPTW. The naive estimate without applying any reweighting method is 0.030 (SE: 0.065). The IPTW effect estimate is -0.006 (SE: 0.068). The TAC effect estimate is 0.002 (SE: 0.079), and the RMT effect estimate is -0.061 (SE: 0.064). All stabilized weights were calculated using the ipw package version 1.2.1 \citep{ipw}. Weights at every measurement point were truncated at the 1st and 99th percentile. Robust standard errors were calculated using the survey package version 4.4-2 \citep{survey}. The naive estimate is slightly optimistic, whereas the regular IPTW estimate and the TAC estimate are both very close to zero. However, the RMT estimate is more pessimistic. Although none of the effect estimates differ significantly from zero, these results indicate that the causal effect estimates from different reweighting methods can differ from one another. Therefore, thought should be put into which reweighting method is most appropriate for the case at hand.

\section{Discussion}
This paper introduces a novel classification of confounding biases that may occur due to irregular measurement times. This categorization is helpful in deciding which reweighting methods to apply in order to get rid of such confounding biases. The categorization extends to both time-as-variable (TAV) graphs and discrete time (DT) graphs. Expert domain knowledge may be used to determine which category of bias is likely to occur in real-life settings. If the measurement time is likely to affect treatment and outcome, there is likely to be direct confounding (DC).  When the possibility of DC is excluded, the visit protocol classification by \cite{pullenayegum_longitudinal_2016} may be useful to distinguish between confounding through measured variables (CMV) and confounding through unmeasured variables (CUV). \textit{History-dependent visits} and \textit{Physician-driven visits} may coincide with CMV, because the measured variables may explain structural differences in measurement time. \textit{Patient-driven visits} likely coincide with CUV, if it is unmeasured factors (i.e. motivation) that act as confounding variables. However, if the researcher is confident to have measured all variables affecting both measurement time and outcome, it may fall under CMV as well. Importantly, the three categories of confounding bias are not mutually exclusive, and may occur at the same time.

In our categorization of backdoor paths it may be observed that the amount of bias is largely dependent on the path from observation time $T$ and observation indicator $N$ to treatment $A$. Furthermore, our simulation results show that the stronger this causal path is, the larger the potential bias may be. This means that the expected severity of this causal path may be used to determine whether it is necessary to adjust for it. In order to identify these situations, we list several example situations in which the association between $T$ and $A$ might be large: 1. When the time since an earlier treatment (strongly) influences future treatment decisions. 2. When the age of a participant has changed significantly over the course of the study, which might directly affect treatment decision. 3. When there is an unmeasured but visible `health' affected by time, which may sway the doctor's decision about treatment for the patient 4. When the time of measurement is correlated with an unmeasured confounder, and acts as a surrogate measure. It is important to keep in mind how strong the path from observation time to treatment is likely to be when assessing whether adjustment for confounding by irregular measurement times is necessary. Inspecting the association between measurement time and treatment decision would give an indication of the strength of this path.

In addition to the categorization, we have compared two ways to denote irregular measurement time into the causal graphs: TAV graphs and DT graphs. The most important way in which these graphs differ from each other is in the definition of the treatment variable. In TAV graphs, the treatment variable denotes the treatment decision at the time of measurement. This means that the treatment variable itself cannot be interpreted separately from the measurement times. As discussed earlier, it is only when estimating the causal effects of specific measurement time-agnostic treatment histories (such as following a treatment regime throughout the entire study period) that the interpretation of the treatment may be separated from the measurement time. We recommend to use TAV graphs only when 1. the researcher is interested in estimating the causal effects of those measurement time-agnostic treatment histories or 2. the researcher is willing to assume that the exact timing of the treatment decision has no impact on the outcome or 3. the researcher can estimate the joint effect of treatment and measurement times on the outcome. 
DT graphs on the other hand are more widely applicable. The treatment variable at each time point is readily interpretable as the potential decision that was made at that time, which naturally allows for an unequal number of measurements per participants, and is not restricted to measurement time-agnostic treatment histories. However, the downsides of this method are that time is discretized, and that TAC may be difficult or impossible to apply.

We have discussed multiple ways to construct the stabilized weights to create a pseudopopulation in which to estimate the ATE. Of the reweighting methods, $\text{SW}^{\text{TAC}}$ was robust to all forms of confounding bias that are part of the categorization. However, the downside of $\text{SW}^{\text{TAC}}$ is that the type of irregular measurement times must be such that it can be described in a TAV graph. As stated earlier, this means that only the ATE of measurement time-agnostic treatment histories may be estimated using $\text{SW}^{\text{TAC}}$. In contrast, $\text{SW}^{\text{RMT}}$ can be applied on both TAV and DT graphs, and was effective at removing the causal effect of measured variables on the time of measurement, as well as selection bias. However, it was not effective at removing bias in the presence of DC and CUV. An additional restriction of $\text{SW}^{\text{RMT}}$ is that it requires discrete time points in order to estimate the probability of being measured at that time, whereas $\text{SW}^{\text{TAC}}$ may be effectively applied in settings where the measurement times are on a continuous scale. We have also found that if there is selection bias as well as confounding bias present, that the use of $\text{SW}^{\text{TAC+RMT}}$ is necessary to adjust for bias. The effectiveness of $\text{SW}^{\text{TAC}}$ and $\text{SW}^{\text{RMT}}$ in settings with an unequal amount of measurements remains to be investigated.

\begin{figure}[h!]
\centering
\includegraphics[width=0.4\linewidth]{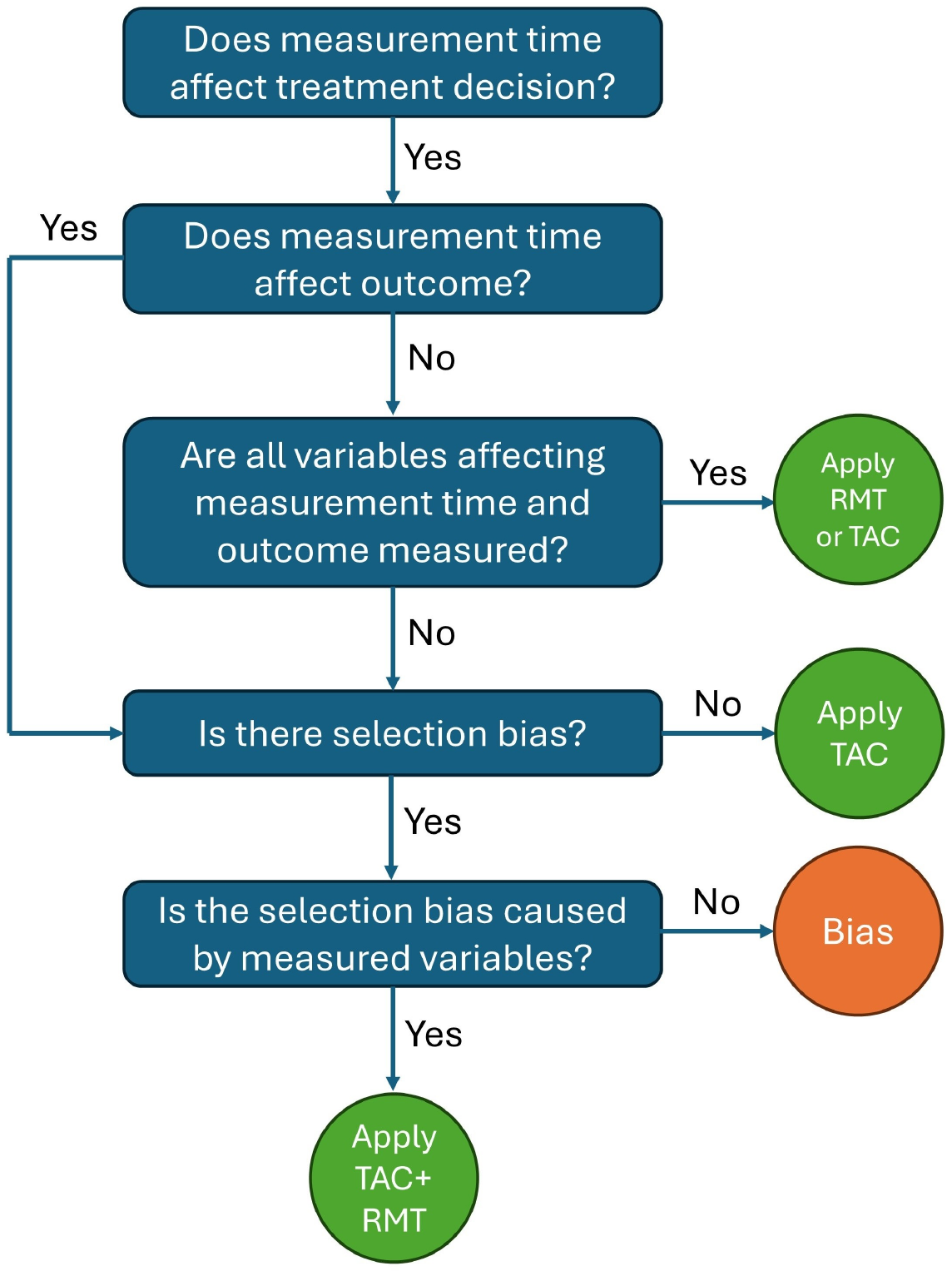}
\caption{\protect\label{fig:Flowchart} A flowchart to guide which reweighting method to apply when using IPTW to estimate treatment effects from data with irregular measurement times. RMT refers to Reweighting by measurement time, TAC refers to Time-as-confounder, and TAC+RMT refers to applying both techniques as described in section 4.4.}
\end{figure}

\subsubsection*{Conclusions}

This study introduced a novel categorization of confounding biases due to irregular measurement times. Furthermore, it was shown that the categorization is important in deciding on the exact reweighting method to use when applying IPTW. This may in turn reduce biases when estimating causal effects from observational data. In practice, researchers may find it hard to determine which category of confounding bias applies, and whether to adjust for it or not. The guidance on which technique to use is summarized in the flow-chart in Figure \ref{fig:Flowchart}. If the goal is to estimate the causal effect of a time-varying treatment that is measurement time-agnostic, and if there is an equal amount of measurements per individual, then we recommend to construct a TAV graph, which allows for the use of $\text{SW}^{\text{TAC}}$ as well as $\text{SW}^{\text{RMT}}$ to adjust for confounding and selection bias. Then, if it is unclear what the exact mechanism of the measurement irregularity was, and therefore which edges might be present in the causal graph, using $\text{SW}^{\text{TAC}}$ is safer than $\text{SW}^{\text{RMT}}$ in terms of preventing confounding bias due to irregular measurement times. Therefore, we recommend to use $\text{SW}^{\text{TAC}}$ when it is unclear what exactly has caused the time irregularity, and $\text{SW}^{\text{RMT}}$ when there is selection bias present. Both reweighting methods may be combined to adjust for as many categories of bias as possible.

\section{Statements and declarations}

\subsection*{\normalsize\bfseries Acknowledgements}
We want to thank Dr. Marieke Hoevenaar-Blom and Dr. Jan Willem van Dalen for their help accessing and on the interpretation of the preDIVA trial data.

\subsection*{\normalsize\bfseries Conflicts of interest}
    The author(s) declared no potential conflicts of interest with respect to the research, authorship, and/or publication of this article.

\subsection*{\normalsize\bfseries Funding}
    This work was supported by a Crossover grant (MOICA 17611) of the Dutch Research Council (NWO). The MOCIA programme is a public-private partnership (see https://mocia.nl/scientific/).

\subsection*{\normalsize\bfseries Data availability}
\noindent The preDIVA data is available from Marieke Hoevenaar-Blom at m.p.hoevenaarblom@amsterdamumc.nl. Access to the data is subject to approval from the steering committee and a data sharing agreement.

\bibliographystyle{apalike}
\bibliography{references}
\section{Appendix}

Here, we describe in more detail the method we used in Section 3.1 to find all backdoor paths that are caused by the irregular measurement times.

\subsubsection*{Constructing the starting graph}

In order to uncover when backdoor paths are opened due to the irregular measurement times $T_1$, we start with example graphs representative of a setting with a time-varying treatment and find all backdoor paths. This graph is depicted in Figure \ref{fig: Procedure1}. For the purpose of staying as general as possible, we want to assume the `most difficult case' in terms of the causal structure of $T_1$, which means that $T_1$ may be affected by any earlier covariates or treatment decisions, and may affect any future covariates, treatment decisions as well as the outcome. However, we remove paths that would violate the base assumptions of IPTW, even if there were no irregular measurement times. The paths that we assume away are:
\begin{itemize}
    \item Causal paths from an unmeasured variable $U$ to $A$ and $Y$, because this would violate the exchangeability assumption of IPTW regardless of the irregular measuerment times
    \item Causal paths from $L$ to $A$. These paths are not of interest to us because they may be adjusted for using regular IPTW.
\end{itemize}

\subsubsection*{Finding all backdoor paths}

The next step of the procedure is to find all of the backdoor paths between treatment history $\bar{A}$ and outcome $Y$. For this procedure we do not condition on any variables, because we will examine bias resulting from conditioning in the next subsection. Given that we do not condition on any variables, it follows that any backdoor path flows towards $\bar{A}$ and $Y$ from a temporally preceding variable, which we call the root of the backdoor path. Thus, any backdoor path has the following conditions:
\begin{itemize}
    \item There must be a downstream path from the root to $Y$
    \item There must be a downstream path from this same root to any treatment decision $A$
\end{itemize}
From each root variable, all possible downstream paths to $A$ and all possible downstream paths to $Y$ were found. Each combination of a downstream path to $A$ and a downstream path to $Y$ formed a backdoor path. An example of a backdoor path consisting of two downstream paths from the root node can be seen in \ref{fig: Procedure2}.

\subsubsection*{Removing redundant parts of the backdoor path}
All of the backdoor paths have been found, but many of the backdoor paths flow through the same variables. If a backdoor path 1 exists that starts at root variable $B$, of which both downstream paths flow through the same variable $C$, then a near-identical path 2 exists that has $C$ as the root variable. A method that corrects for the causal path from $C$ towards treatment and outcome will simultaneously adjust for both path 1 and 2. Therefore, these paths are not interesting to examine separately. As an example, see Figure \ref{fig: Procedure3}. In order to restrict the backdoor paths to the relevant backdoor paths, the following condition was added:
\begin{itemize}
    \item The downstream paths do not flow through the same variable, except for the root
\end{itemize}
Thus, to find all relevant backdoor paths, all backdoor paths where the path to $A$ and the path to $Y$ flowed through the same variable were removed. A list of all the resulting relevant backdoor paths can be found in the supplementary material.
 
\end{document}